\RequirePackage{fix-cm}
\documentclass[twocolumn]{svjour3}          
\smartqed  
\usepackage{graphicx}
\usepackage{subfigure}
\usepackage{bm}
\usepackage{amsmath,ulem}
\usepackage{amssymb}
\usepackage{mathptmx}      
\usepackage{latexsym}
\usepackage{verbatim}
\usepackage{color}
\usepackage[english]{babel}
\usepackage{epstopdf}
\usepackage{amsmath}
\usepackage{cite}

\newcommand{\bs}[1]{\boldsymbol{#1}}
\journalname{Journal of Computational Electronics}
\begin{document}
\title{Atomistic deconstruction of current flow in graphene based hetero-junctions}


\titlerunning{Atomistic Non-equilibrium Green's function simulation of graphene based heterojunctions}        

\author{Redwan N. Sajjad \and
        Carlos Polanco \and        
        Avik W. Ghosh
}


\institute{Redwan N. Sajjad \and
           Carlos Polanco \and
Avik W. Ghosh  \at
           Department of Electrical and Computer Engineering, Charlottesville, VA-22904, USA  \\
              \email{redwan@virginia.edu}           
           }

\date{Received: date / Accepted: date}

\maketitle

\begin{abstract}
We describe the numerical modeling of current flow in graphene heterojunctions, within the Keldysh Landauer Non-equilibrium Green's function (NEGF) formalism. By implementing a $k$-space approach along the transverse
modes, coupled with partial matrix inversion using the Recursive Green's function Algorithm (RGFA), we can simulate on an atomistic scale current flow across devices approaching  experimental dimensions. We use the numerical platform to deconstruct current flow in graphene, compare with experimental results on conductance, conductivity and quantum Hall, and deconstruct the physics of electron `optics' and pseudospintronics in graphene $p-n$ junctions. We also demonstrate how to impose exact open boundary conditions along the edges to minimize spurious edge reflections. 

\end{abstract}

\section{Introduction}\label{sec:intro}
Ever since its inception \cite{novoselov_04}, graphene as a material has occupied a unique place
between one dimensional pi-conjugated organic molecules {and} three dimensional
bulk crystalline solids with
well defined bandstructures. The quasi-ballistic scattering lengths \cite{morozov_08,bolotin_08}, photon like 
dispersion \cite{weiss_58} and chiral electron flow \cite{katsnelson_06} together promote nontrivial transport physics in graphene, 
opening up various device possibilities that necessitate detailed numerical 
modeling. While the photonic bandstructures argue for a continuum model of modest 
sized device segments, scattering at heterojunction interfaces \cite{ref1} and edges requires  
a detailed atomistic treatment. The aim of this paper is to describe practical techniques for 
simulating quantum transport through graphene based heterojunctions, compare 
various transport regimes with experiments \textcolor{black}{associated with the chiral nature of electron transmission at graphene heterojunctions}. We also point out some of the unique electronic properties of bilayer graphene, but unless otherwise stated, the results are for single layer graphene.  
\textcolor{black}{\section{Simulation platform}}

The Non-equilibrium Green's function (NEGF) formalism \cite{datta_97} provides a unified, `bottom-up'
platform for modeling quantum flow of electrons. Indeed, it has been successfully applied
to understand transport physics in materials and systems as diverse as organic molecules \cite{damle_01}, carbon nanotubes \cite{guo_04,koswatta_07}, graphene \cite{yoon_07, tseng_09, low_09, low_11, sajjad_11, sajjad_12}, silicon nanowires \cite{wang_04, bescond_04, luisier_06, nehari_07, shin_07, sajjad_09_param}, spintronics, nanomagnets \cite{rocha_05,sayeef_06}, nanoscale phonon transport \cite{mingo_03, mingo_06, wang_08}. A challenge however is the sheer problem size associated with atomistic deconstruction of experimentally relevant dimensions, typically hundreds of nanometers, as well as the scattering physics at the atomistically rough edges. We employ a recursive technique to simplify the problem size in order to make such a deconstruction tractable.

{\bf{\subsection{Recursive Green's function Algorithm (RGFA) }}}
The central entity for quantum kinetics in a weakly
interacting system is the retarded Green's function, defined as,
\begin{eqnarray}\label{gr}
\mathcal{G} = (EI-H-\Sigma_1-\Sigma_2)^{-1}
\end{eqnarray} $H$ is the Hamiltonian matrix of graphene, described in this paper using a minimal one $p_z$ orbital basis per carbon atom \textcolor{black}{with $t_0 = -3$eV being the hopping parameter.} $\Sigma_{1,2}$ are the self energy matrices from the source and drain contacts, and $\Gamma_{1,2}$ are the corresponding anti-Hermitian parts representing the energy level broadening  associated with charge injection and removal. The corresponding spectral function and quantum mechanical total transmission are given by 
\begin{eqnarray}
A &=& i(\mathcal{G}-\mathcal{G}^\dagger)\nonumber\\
T &=& Tr(\Gamma_1\mathcal{G}\Gamma_2\mathcal{G}^\dagger)
\end{eqnarray}
{{\textcolor{black}{The local density of states is obtained from the diagonal elements of $A$ divided by $2\pi$}}}.
 
{{\textcolor{black}{The contact self energy matrices $\Sigma_{1,2}$ are calculated from, $\Sigma =\tau g \tau^\dagger$, where $g$ is the surface Green's function of the contact and $\tau$ is the coupling matrix between the contact and graphene. The surface Green's function $g$ is found by solving the following equation recursively 
\begin{eqnarray}
g&=&[\alpha-\tau_1 g \tau_1^\dagger]^{-1}
\end{eqnarray}
where $\alpha = EI-H$ for the layered  contact unit cell.}}} For speedy convergence of $g$, we use a decimation technique outlined in \cite{galperin_02}.

Inverting the matrix in Eq. \ref{gr} is computationally expensive and goes out of hand even for a small structure (width$\textgreater$5nm). The Recursive Green's Function Algorithm (RGFA) is a fast practical method to avoid inverting the entire matrix by brute force, but instead extract only select, relevant blocks of the $\mathcal{G}$ matrix, as described in Ref. \cite{anantram_02}, \cite{lake_97}. For
instance the transmission across a layered device involves only a corner block of the $\mathcal{G}$ matrix, $\mathcal{G}_{1,L}$, while the density of states involves only a diagonal block $\mathcal{G}_{L,L}$,  enabling 
such partial inversion. 
\begin{figure}
\centering
\includegraphics[width=3in]{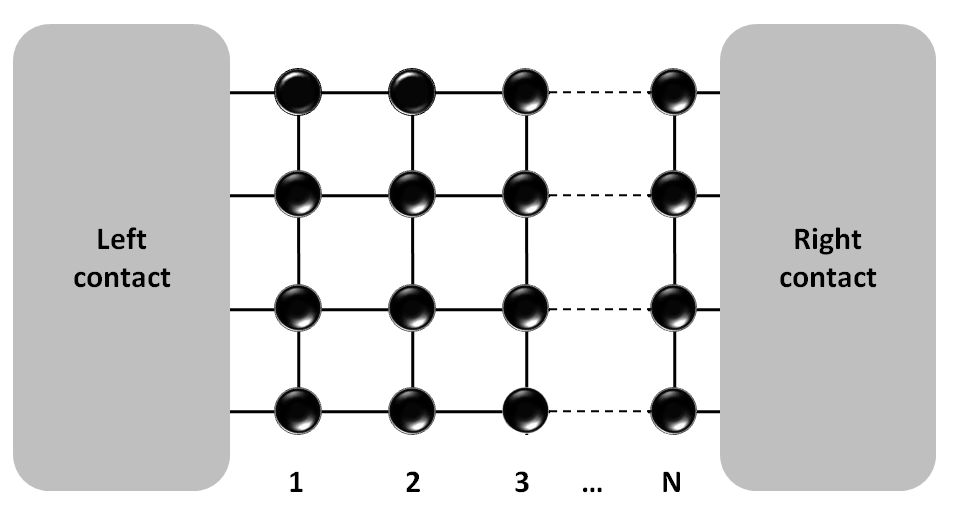}\quad
\caption{Recursive Green's Function Algorithm (RGFA) requires partitioning the structure in left contact, device (layers of N) and right contact.}
\label{layers}
\end{figure}
{\bf{\subsubsection{Evaluating the Green's functions piece by piece}}}
We consider the device area in layers of N as shown in Fig. \ref{layers}. The right connected Green's function \cite{lake_97} is calculated from
\begin{eqnarray}
g_{L,L} = (EI-H_L-U_L-t_{L,L+1}+g_{L+1,L+1}t_{L+1,L})^{-1}
\label{gll}
\end{eqnarray} where $H_L$ is the Hamiltonian of the $L$ th layer, \textcolor{black}{ $t_{L,L+1}$ is the coupling matrix between adjacent layers and $U_L$ is the electrostatic potential of the $L$th layer}. The calculation of $g_{L,L}$ {{\textcolor{black}{is done by stepping through the above equation from right to left starting}}} at layer $N-1$, {{\textcolor{black}{with the right connected Green's function}}} for the $N$th layer\\
\begin{eqnarray}
g_{N,N} = [EI-H_L-U_L-\Sigma_2]^{-1}
\end{eqnarray}{{\textcolor{black}{Once we reach layer 1 and extract $g_{2,2}$, the device Green's function component for that layer is then calculated  from}}}
\begin{eqnarray}
\mathcal{G}_{1,1} = (EI-D_1-\Sigma_1-t_{1,2}g_{2,2}t_{2,1})^{-1}
\end{eqnarray} where $D_1 = H_{1,1}+U_1$. The remaining layer block diagonal matrices are calculated {{\textcolor{black}{by stepping through from left to right  using,}}}
\begin{eqnarray}
\mathcal{G}_{L,L} = g_{L,L}+g_{L,L}t_{L,L-1}\mathcal{G}_{L-1,L-1}t_{L-1,L}g_{L,L}
\end{eqnarray} with $L = (2,\ldots, N)$. 
\bigskip
\bigskip
{\bf{\subsubsection{From Green's function to conductance and non-equilibrium carrier density}}}
From the {{\textcolor{black}{diagonal blocks of the layer}}} Green's functions, we can calculate the spectral function 
\begin{eqnarray}
A_{L,L} = i(\mathcal{G}_{L,L}-\mathcal{G}_{L,L}^{\dagger})
\end{eqnarray}The left connected spectral function for layer $L$ is given by,
\begin{eqnarray}
A_{L,L}^L = \mathcal{G}_{L,1}\Gamma_{1,1}\mathcal{G}_{L,1}^{\dagger}
\end{eqnarray}
{{\textcolor{black}{which requires evaluating a corner block of $\mathcal{G}$ }}}
\begin{eqnarray}\label{gl}
\mathcal{G}_{L,1} = g_{L,L}(-t_{L,L-1})\mathcal{G}_{L-1,1}
\end{eqnarray}
{{\textcolor{black}{once again by stepping through from left to right, starting with $\mathcal{G}_{1,1}$ in layer L=2}}}.
All the matrices from {{\textcolor{black}{Eqs.~\ref{gll}-\ref{gl}}}} are $N_L \times N_L$, where $N_L$ is the number of atoms in one layer. Since matrix inversion is an $N^3$ procedure, inverting substantially smaller matrices at a time leads to considerable computational savings overall. 

The total conductance  is calculated from the transmission over all modes
\begin{eqnarray}\label{negf_t}
G = G_0\cdot Tr(\Gamma_{1,1}[A_{1,1}-\mathcal{G}_{1,1}\Gamma_{1,1}\mathcal{G}_{1,1}^{\dagger}])
\end{eqnarray} 
{{\textcolor{black}{where $G_0 = 2q^2/h$ takes care of spin degeneracy, while valley degeneracy is captured with the graphene Hamiltonian}}}

The non-equilibrium carrier density at each layer is calculated from the above quantities as
\begin{eqnarray}
\rho_L = 2\int_{-\infty}^{\infty}\frac{dE}{2\pi}Tr[{f_1A_{L,L}^L+f_2(A_{L,L}-A_{L,L}^L)]}
\end{eqnarray}where $f_1$ and $f_2$ are Fermi functions at the source and drain respectively.
\bigskip
\bigskip
{\bf{\subsubsection{Extracting current density}}}
The current from $i$th atom to $jt$h atom is calculated from \cite{datta_97},
\begin{eqnarray}\label{crnt}
I_{i,j} = \frac{2q}{h}\int dE Im[\mathcal{G}^n_{i,j}(E)H_{j,i}-H_{i,j}\mathcal{G}^n_{j,i}(E)]
\end{eqnarray}where the electron correlation function, $\mathcal{G}^n=\mathcal{G}\Sigma^{in}\mathcal{G}^{\dagger}$ and scattering function, $\Sigma^{in} = \Gamma_Sf_S+\Gamma_Df_D$. The source and drain Fermi levels are at $\mu_S=0$ and $\mu_D = -qV_D$. To see the current distribution in the device, we apply a small drain bias $V_D$. $I_{i,j}$ is nonzero only if  the $i$th atom and $jt$h atom are neighbors. The total current at one atomic site can be found by adding all the components vectorially, \textcolor{black}{$I_i = |\sum_jI_{i,j}e^{i\Phi}|$, where $\Phi$ is the angle of the individual current vectors with respect to a reference direction. As expected, the NEGF expression for $I_{i,j}$ satisfies the steady-state Kirchhoff's law, $\sum_{i}I_{i} = 0$.} The terminal current $I_T$ is the sum of all currents in one atomic layer, and should be equal to the current from the Landauer formalism using the total transmission,
\begin{eqnarray}\label{tc}
\textcolor{black}{I_{T} = \int_{-\infty}^{\infty} \displaystyle \frac{G(E)}{q}[f_S(E)-f_D(E)]dE}
\label{curr}
\end{eqnarray}
{{\textcolor{black}{While the total current $I_T$ (Eq.~\ref{curr}) can be calculated efficiently using the partial blocks in the RGFA, the spatially resolved current density (Eq. \ref{crnt}), useful primarily for visualization purposes, employs the full matrix $\mathcal{G}$ and is thus computationally expensive. }}}
The recursive algorithm 
to calculate blocks of $\mathcal{G}^n$ is described in \cite{anantram_02}. The current density from layer $L$ to $L+1$ is given by,
\begin{eqnarray}\label{current_den}
I_{L\rightarrow L+1}(E) = \frac{2q}{h}Im[t_{L,L+1}\mathcal{G}^n_{L,L+1}(E)-t_{L+1,L}\mathcal{G}^n_{L+1,L}(E)]
\end{eqnarray}The matrix $I_{L\rightarrow L+1}(E)$ has the size of $N_L\times N_L$ carrying the atom to atom current components between the two layers. \\

{\bf{\subsection{$k$ space formalism (KSF)}}}
We can further cut down the problem size when the system is periodic along the direction perpendicular to electron transport, by employing Bloch's theorem. The periodicity allows us to decompose the system so that the transverse k-points act as decoupled 1-D chains or modes, whose contribution to the total current can be calculated independently. We can thereafter calculate the conductance through bulk graphene or graphene nanoribbons (GNRs) with mode by mode resolution. As we shall shortly see, it also allows us to  visualize the transition from the smooth transmission profile of bulk graphene to the stepwise transmission of GNRs driven by width quantization.

\begin{figure}[tb]
\centering
\includegraphics[width=86mm]{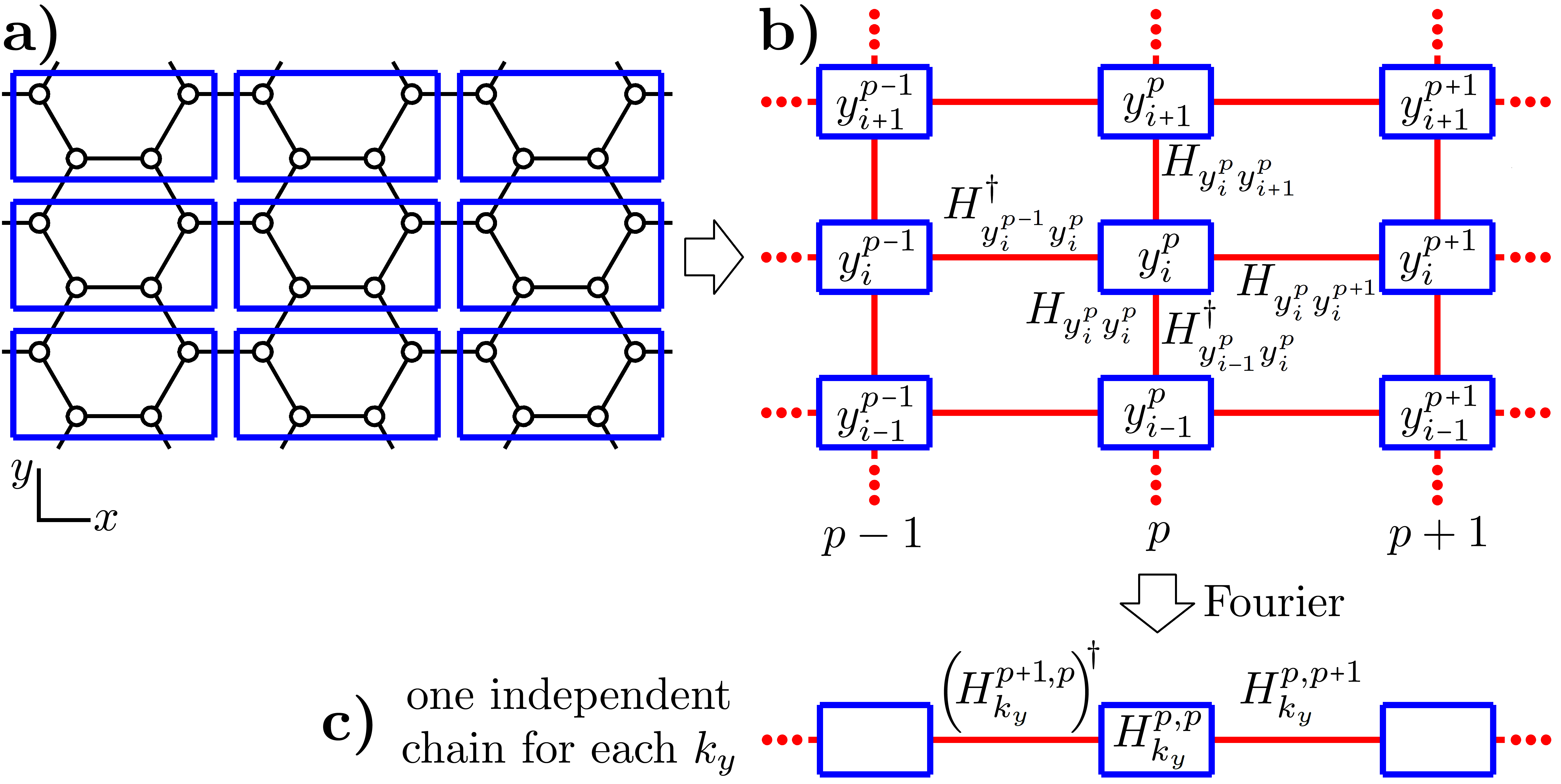}
\caption{{\bf a)} Graphene sheet split into blocks that reflect its periodicity in the $y$ direction. {\bf b)} Representation of Graphene's hamiltonian split into matrices describing the interaction between block's wave functions. Each box represents the interaction of the a block with itself and each line represents the interaction of a block with a neighboring block. {\bf c)} After Fourier transforming the Graphene sheet can be split into independent chains, one for each $k_y$.}
\label{figGraphene}
\end{figure}

Consider a graphene sheet with an applied potential varying only in the $x$ direction (Fig.~\ref{figGraphene}{a}). After dividing the sheet into blocks, we can split the hamiltonian into $4\times4$ \textcolor{black}{block} matrices ($H_{y_i^py_j^q}$) describing the interaction between the wave functions $\bs{\psi}_{y_i^p}$ and $\bs{\psi}_{y_j^q}$ belonging to blocks  $y_i^p$ and $y_j^q$, \textcolor{black}{with subscripts  denoting the rows labeled along the transverse direction, and superscripts denoting layer columns labeled along the transport direction} (Fig.~\ref{figGraphene}{\bf b}). Using this pictorial representation of the Hamiltonian, the Schr\"odinger equation for the $y_i^p$ block is given by
\begin{align}
E\bs{\psi}_{y_i^p}=&H_{y_i^py_i^p}\bs{\psi}_{y_i^p}+H_{y_i^py_{i+1}^p}\bs{\psi}_{y_{i+1}^p}+H_{y_{i-1}^py_i^p}^\dagger\bs{\psi}_{y_{i-1}^p} \nonumber \\
&+H_{y_i^py_i^{p+1}}\bs{\psi}_{y_i^{p+1}}+H_{y_i^{p-1}y_i^p}^\dagger\bs{\psi}_{y_i^{p+1}}. \label{equSEnp}
\end{align}
To take advantage of the manifest periodicity in $y$, we assume Bloch type solutions  $\bs{\psi}_{y_j^q}=\tilde{\bs{\psi}}_pe^{i\bs{k}_y\cdot \bs{y}_j^q}$, with $\bs{y}_j^q$ the vector describing the position of the block $y_j^q$. Substituting this solution into Eq.~\ref{equSEnp}, we get 
\begin{equation}
E\tilde{\bs{\psi}}_p=H^{p,p}_{k_y}\tilde{\bs{\psi}}_p+H^{p,p+1}_{k_y}\tilde{\bs{\psi}}_{p+1}+\left(H^{p-1,p}_{k_y}\right)^\dagger\tilde{\bs{\psi}}_{p-1},
\label{equSEpkspace}
\end{equation}
with
\begin{equation}
H^{p,p}_{k_y}=H_{y_i^py_i^p}+H_{y_i^py_{i+1}^p}e^{ik_yb}+H_{y_{i-1}^py_i^p}^\dagger e^{-ik_yb},
\end{equation}
\begin{equation}
H^{p,p+1}_{k_y}=H_{y_i^py_i^{p+1}},
\end{equation}
\begin{equation}
H^{p-1,p}_{k_y}=H_{y_i^{p-1}y_i^p}.
\end{equation}
For each $k_y$, Eq.~\ref{equSEpkspace} with a varying index $p$ can be interpreted as a one dimensional chain (Fig.~\ref{figGraphene}{\bf c}) decoupled from chains with other $k_y$s. In other words, we have split the graphene sheet into a set of independent chains or modes. Now, we can use the usual NEGF formalism to find the transmission of each chain or mode ($T_{k_y}$).
\begin{equation}\label{emode}
T_{ky}=Tr(\Gamma^1_{k_y}\mathcal{G}_{k_y}\Gamma^2_{k_y}\mathcal{G}_{k_y}^\dagger)
\end{equation}
The quantities ($\Sigma$, $\Gamma$, $G$ and $T$) in real space are found by inverse Fourier transform, since our Bloch assumption is equivalent to Fourier transforming from the discrete $y_i$-space to $k_y$-space. For example, the Green's function is given by
\begin{equation}
\mathcal{G}(y_i^p,y_j^q)=\frac{1}{N_k}\sum_{k_y}\mathcal{G}_{k_y}e^{i\bs{k}_y\cdot(\bs{y}_j^q-\bs{y}_i^p)},
\end{equation} with $N_k$, the number of points in $k_y$ space, and the $k_y$ dependent Green's function given by

\begin{eqnarray}\label{gk}
\mathcal{G}_{k_y}=[(E+i\eta)I-H_{k_y}-\Sigma_{1,k_y}-\Sigma_{2,k_y}]^{-1}
\end{eqnarray}
The sigma matrices are calculated as before, $\Sigma_{k_y}=\tau_{1}^\dagger g_{k_y} \tau_{1}$ but the surface Green's function is now $k_y$ dependent.
\begin{eqnarray}
g_{k_y}&=&[\alpha_{k_y}-\tau_1 g_{k_y} \tau_1^\dagger]^{-1}
\end{eqnarray}where $\alpha_{k_y}=EI-H_{k_y}$. $H_{k_y}$ takes care of all the couplings in the transverse direction.

Thus the combination of semi-infinite contacts in the $\textcolor{black}{\bf{\pm}} x$ direction and Fourier transformation along the {\bf{\textcolor{black}{$\pm$}}}$y$ direction amounts to a change in basis set for the completely periodic bulk 2D graphene.
Since trace is preserved under the basis transformation, we can calculate transmission directly in 
$k_y$ space 
\begin{equation}
T=\sum_{k_y}T_{k_y}\\
DOS = \frac{i}{2\pi}Tr(\mathcal{G}-\mathcal{G}^{\dagger})
\end{equation}where
\begin{eqnarray}\label{g}
\mathcal{G}= \sum_k{\mathcal{G}_{k_y}}
\end{eqnarray}\\
The distinction between bulk graphene and GNR arises primarily from the different sets of $k_y$'s (in practice, one needs to worry about edge reconstruction, next nearest neighbor interactions, strain and roughness at the edges of GNRs. For an ideal GNR of width $W=Nb=N_kb$ the hard wall boundary conditions restrict $k_y$ to the discrete set\\\\
$k_y=\displaystyle \frac{n\pi}{W} \text{ with } 0\leq k_yb \leq \pi \text{ and } n \text{ integer}.$\\\\
In the case of bulk graphene we can assume periodic boundary conditions every $W=Nb$, which restrict the $k_y$'s to 
\\\\$k_y=\displaystyle\frac{2n\pi}{W} \text{ with } -\pi \leq k_yb \leq \pi \text{ and } n \text{ integer},$\\\\
and let $N\rightarrow\infty$. Note that as $N$ increases, the transmission evolves from a stepwise curve showing width quantization to a smooth curve.\\


{\bf{\subsection{Applying open boundary conditions}}}

Since simulations deal with finite sized matrices, it is important to make sure the results
do not get influenced by edge effects. While realistic physical devices do have edges, their 
impact on transport is complicated, dominated by strain and reconstruction, roughness, 
possible charge trapping, edge dipoles, localized vibrons and other subtleties. In fact, measurements of
tilt-dependent junction resistance in graphene show a surprisingly weak influence of
edge scattering \cite{sajjad_12,sutar_12}. For modeling purposes therefore,
it is important to ensure that additional edge effects do not creep in simply 
because of the finite-sized domain of our simulation grid. 

Section 2.1 deals with hard-wall boundary conditions along the transverse direction.
For such structures, eliminating edge effects requires making the widths very long
(often longer than device length), which may have added repercussions on material
properties (see section 3.1). Section 2.2 implemented periodic boundary conditions (PBC)
instead with a k-space formalism, allowing us to work with just a couple of nearest 
neighbor unit cells along the transverse direction, but
requiring strict periodicity along that direction. In this section, we will 
discuss how to implement open boundary conditions (OBC) that would allow us to
irreversibly remove an electron impingent at the edges. The trick is to treat the edges
as virtual `contacts' with self-energy matrices built out of complex numbers, acting 
like energy-dependent lossy potentials.

\begin{figure*}
\centering
\includegraphics[width=5in]{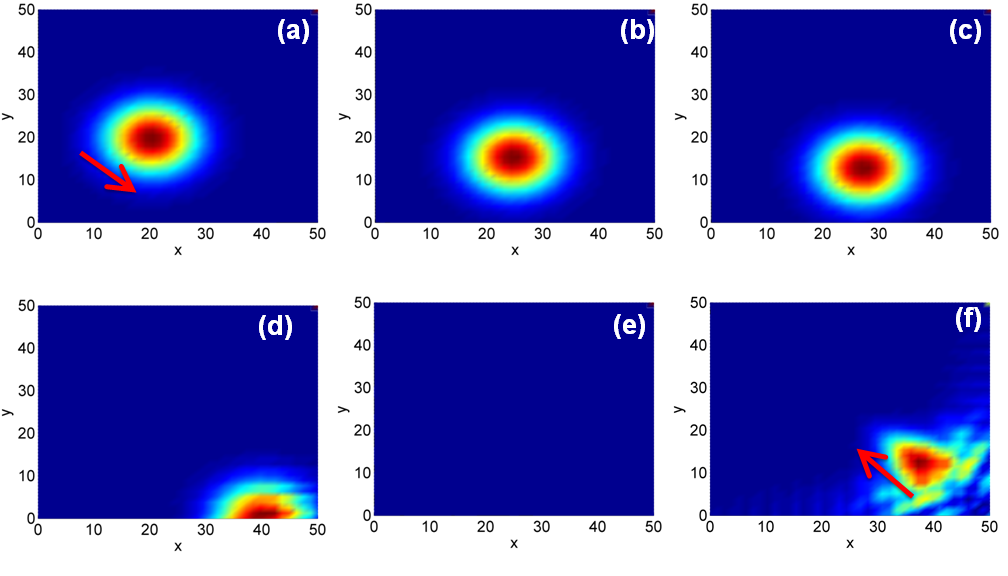}\\
\subfigure{\includegraphics[width=3.8in]{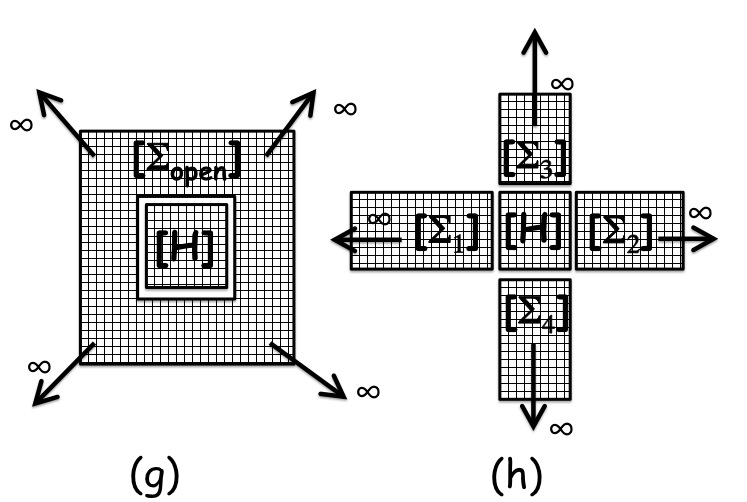}}
\caption{Tracking the injected Gaussian wavepacket at different times (a-e), the wavepacket escapes due to the {exact open boundary condition (g) imposed through the self energy term. We see some reflections along the corner if we adopt a less exact (but computationally less expensive) method to impose open boundary condition through transverse contacts (h).}}
\label{open}
\end{figure*} 

In this section, we will show how to generate exact open boundary conditions (OBC) at the 
edges of a device. It is worth emphasizing at this point that these are {\it{quantum boundary
conditions}}, i.e., acting on the retarded Green's function $\mathcal{G}$. What is much harder to solve is a {\it{thermal open boundary condition}} acting on $\mathcal{G}^n$, because this 
depends on details of the scattering processes outside the simulation regime. In metallurgical
contacts, we impose by fiat Fermi-Dirac distributions, but in the extended geometry just 
below our simulation regime, the distribution is non-equilibrium and the corresponding
boundary conditions need to be estimated {\it{self-consistently}} with the $\mathcal{G}^n$ inside the device in order to prevent the outside regime from sucking out the electrons too
aggressively or else inadequately. \\

{\bf{\subsubsection{The `scooped' open boundary self-energy}}}
The `obvious' way to implement some kind of open boundary is to attach virtual
leads along the transverse direction and extract their self-energies $\Sigma_{3,4}$
using recursion. This is schematically described by Fig.~\ref{open}(h). While this 
method will take the electrons away, the leads do not completely span the outside
regime relative to the central device represented by the Hamiltonian [H], and the
missing chunks at the corners are expected to create spurious reflections. One
could extend the contacts $\Sigma_{1,2}$ or $\Sigma_{3,4}$ laterally using $k$
space formalism to span those corner regions, but the problem is that the four virtual
leads are ultimately decoupled from each other with no bonds running in between,
and thus do not quite represent a straightforward extension of the device domain beyond 
the simulation regime. 

What a proper OBC will need is a geometry like Fig.~\ref{open}(g), where the virtual 
`contact' should be a single monolithic structure with the 
central device region `scooped' out. The single self-energy matrix $\Sigma_{open}$ for this
outside regime must be reverse engineered so that 
the Green's function of the central region matches that of the infinite system periodically
extended along all axes.  Once we find this self energy (section 2.3.2), we can use it for
any modified condition (e.g. a gate voltage, a molecular adsorbate or an injecting contact) 
that only influences the central region but keeps the outside unaltered. \\
{\bf{\subsubsection{Calculating self energy $\Sigma_{open}$}}}
To find the boundary free real-space Green's function in the central region, we first extract
the $k$ space version corresponding to the infinite periodic system.
\textcolor{black}
{\begin{eqnarray}
\mathcal{G}_{\vec{k}}=[(E+i\eta)I-H_{\vec{k}}-\Sigma_{1,\vec{k}}-\Sigma_{2,\vec{k}}]^{-1}
\end{eqnarray}}
\noindent where the k-dependent quantities are Fourier transformed sums over nearest neighbors. 
From this we can calculate the real-space Green's function projected onto the device
region by inverse transforming 
\begin{equation}
\mathcal{G}_{mn} = \sum_{\vec{k}}\mathcal{G}_{\vec{k}}\exp{[-i\vec{k}\cdot(\vec{R}_m-\vec{R}_n)}]/N_mN_n
\end{equation}
 for the atoms `m' and `n' spanning
the central device segment. For a system with $N$ atoms, $\mathcal{G}$ is $N \times N$. 

We can now reverse engineer the open boundary self-energy, armed with this real space
Green's function and the $N \times N$ device Hamiltonian $H$, giving us
\begin{equation}
\Sigma_{open} = EI - H - \mathcal{G}^{-1}
\end{equation} This self energy can be used to calculate transmission through the structure with open boundary conditions, additional injecting/removing leads $\Sigma_{1,2}$ and any other perturbations acting only on the central region through a localized potential $V$ that does not influence the boundaries
\begin{eqnarray}
T &=& Trace(\Gamma_S \mathcal{G}_{new} \Gamma_D \mathcal{G}_{new}^{\dagger})\nonumber\\
\mathcal{G}_{new} &=&  (EI-H-V-\Sigma_{S}-\Sigma_{D}-\Sigma_{open})^{-1}
\end{eqnarray}\\

{\bf{\subsubsection{Verifying the open boundary condition}}}
In Fig.~\ref{open}, we verify the efficiency of the open boundary condition by launching a Gaussian wavevector
with a spread $\{\sigma_x,\sigma_y\}$ and initial quasi-momentum set by the wave-vector $\{k_x,k_y\}$,
\begin{eqnarray}
\Psi(x,y,t=0) = \frac{1}{2\pi{\bf{\textcolor{black}{\sqrt{\sigma_x\sigma_y}}}}}&e&^{-(x-x_0)^2/2\sigma_x^2}e^{-(y-y_0)^2/2\sigma_y^2}\nonumber\\
\times &e&^{-ik_xx}e^{-ik_yy}
\end{eqnarray} about an injection point $(x_0,y_0)$. We calculate the evolution of the gaussian wave-packet with time by numerically solving the time-dependent Schr\"odinger equation within the 
Crank Nicholson algorithm (described below), acting on the time-dependent vector $\{\Psi(t)\}$ with row entries
corresponding to the spatial coordinates
\begin{eqnarray}
&&\{\Psi(t+\Delta t)\} = \bigl[M\bigr]\{\Psi(t)\}\nonumber\\
&&M = \Biggl[I - i\biggl(H+\Sigma_{open}\biggr)\Delta t/2\Biggr]\Biggl[I + i\bigl(H+\Sigma_{open}\bigr)\Delta t/2\Biggr]^{-1}\nonumber\\
&&
\end{eqnarray}
In the absence of any open boundaries, the wavepacket stays confined within the simulation
domain and bounces around (not shown). With the scooped self-energy $\Sigma_{open}$ (Fig.~\ref{open}g), the
open boundary condition takes the electron out at the boundary (Figs.~\ref{open}a-e). Note
that the self-energy we used works only at a specific energy $E$ set by the pair $(k_x,k_y)$, although we can extend it to all energies by Fourier transforming.

Fig.~\ref{open}(f) shows the results for a cross-geometry with transverse contacts represented by Fig.~\ref{open}(h). As expected, the missing pieces along the corners give a sizeable reflection of the wavepacket. \\

\section{NEGF simulation of electron transport in graphene}
In the following sections, we will use a combination of numerical techniques outlined so far to simulate current flow through graphene segments comparable in size to experimental dimensions. We show how these simulations accurately capture the nuances of conductivity,
conductance, magnetotransport, pseudospintronics and Klein tunneling within a common, unified simulation platform. 

{\bf{\subsection{Minimum conductivity vs conductance}}} One of the unique properties of graphene is that its \textcolor{black}{minimum} {\it{conductivity}} $\sigma$ (rather than conductance $G = \sigma W/L$) is quantized in units of $4q^2/\pi h$, when the width to length aspect ratio $W/L$ is large. In other words, $G$ is proportiontal to the aspect ratio $W/L$ in wide graphene samples. The width-dependence is expected in a large ballistic device as the number of modes is roughly  proportional to the number of Fermi half-wavelengths one can fit into $W$. However, the average transmission per mode, and thus the conductance, is usually length-independent. This is because propagating modes have transmissions of unity in a ballistic channel, while evanescent tunneling modes have transmissions that are nearly zero. However since graphene is semi-metallic, a wide sheet supports a nearly continuum set of sub-bands with ultralow band-gaps and tunneling probabilities that are not ignorable. In fact the $\sim 1/\cosh^2 q_yL$ tunnel transmission terms ($q_y = n\pi/W$) from these closely spaced modes  all add up to an overall $1/L$ dependence of the total transmission (more precisely, a factor $2W/L\pi$) and therefore the conductance $G$, leading to the conductivity quantization. 
Experiments on dirty samples show such a quantized minimum conductivity, albeit off by a 
factor of $\sim 3-4$, usually attributed to Coulomb scattering from charge puddles \cite{tan_07, miao_07, chen_08}.

\begin{figure}
\centering
\includegraphics[width=2.8in]{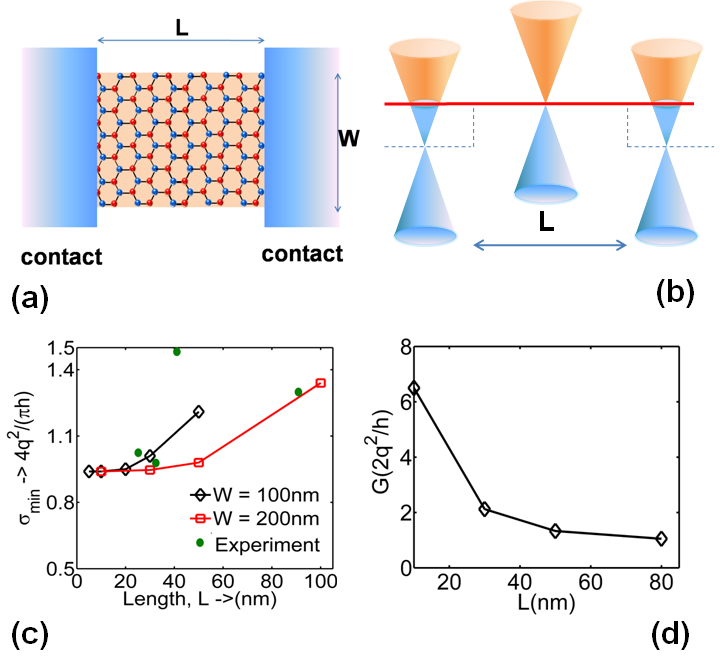}\quad
\caption{Numerical extraction of graphene minimum conductivity in the small dimension limit (L\textless500nm). (a) Schematic of the device, (b) Energy band diagram for the device with doped contacts, (c) Calculated $\sigma_{min}$  and (d) $G_{min}$for various aspect ratios along with experimental data from Ref. \cite{miao_07}.}
\label{graphene}
\end{figure}

We will use the NEGF technique to extract the conductivity of graphene at the Dirac point. To recap, for large scale graphene devices ($L>1\mu m$), the {{\it{conductance per mode}}} is expected to be  $ 4q^2/h$ (the mode count is given by the integer part of $Wk_F/\pi$), while for smaller lengths ($L<500nm$) and larger widths, the {\it{conductivity}} is expected to approach $4q^2/(\pi h)$. We will simulate the structure shown in Fig. \ref {graphene}(a) to calculate the minimum conductivity of a clean graphene sheet. The graphene device is connected to two heavily doped graphene contacts. As long as the contacts provide a large number of modes and thus minimal series resistance, the details of the contact metallicity should not matter. The energy band diagram for the problem is shown in Fig. \ref {graphene}(b). We model the system atomistically \textcolor{black}{with RGFA as described in section 2.1.}

We use Eq. \ref{negf_t} to calculate the total conductance $G$.  Minimum conductivity $\sigma_{min} = G{L}/{W}$ is shown in Fig. \ref{graphene}(c) along with the experimental data from Ref. \cite{miao_07}. We take two different widths $W$ = 100, 200 nm to calculate $G$ and $\sigma_{min}$ for various lengths ($L$). $\sigma_{min}$ is found to be very close to ${4q^2}/{\pi h}$. As expected, the minimum conductivity stays close to this number as long as $W/L$ is large and a continuum evaluation of the tunnel contribution is a valid approach. In the opposite limit $W \ll L$ the conductivity increases with length and we reach the conventional regime of conductance quantization (Fig.~\ref{graphene}(d)). \textcolor{black}{Note that for a graphene ribbon of finite width, the valley degeneracy of the fundamental mode is broken \cite{twor_06}, so that the minimum conductance is $2q^2/h$.} At higher gate voltages as more modes come into play, the conductance per mode reaches $4q^2/h$, with the degeneracy of 4 coming from the presence of 2 spins and 2 valleys at the same energy. 

\begin{figure}
\centering
\includegraphics[width=3.3in]{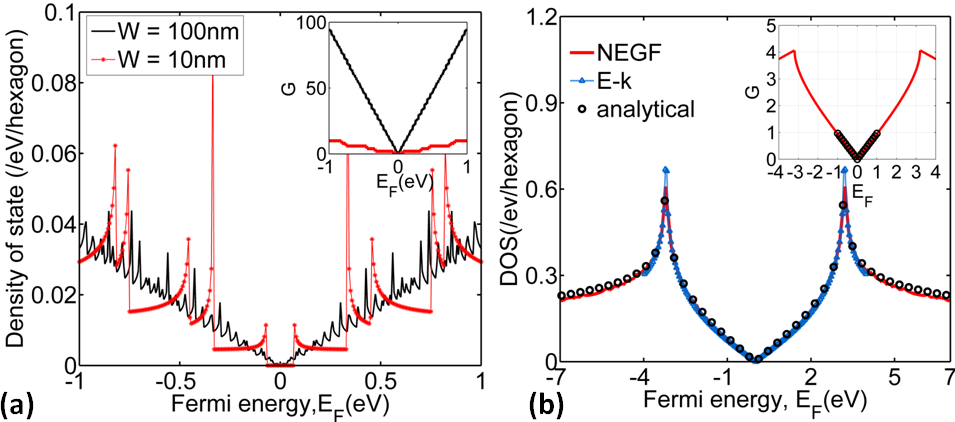}\quad
\caption{(a) Density of states (DOS) and conductance (inset, in units of $2q^2/h$) calculation from atomistic tight binding RGFA for two different widths, (b) Similar calculations over a wider energy range done with KSF, atomistic NEGF within RGFA, and an analytical integration of the $E-k$ dispersion. Conductance (inset) is in units of $2q^2/h\times1000 \mu m^{-1}$. Conductance from KSF (red line) agrees with simple linear approximation of no. of modes (black circles) at low energy.}
\label{gnr}
\end{figure}

{\bf{\subsection{RGFA for finite sized graphene devices}}}
We will now apply the RGFA technique discussed in section II to simulate three types of finite sized graphene based devices - uniformly doped, $pn$ \textcolor{black}{step} junction and $npn$ \textcolor{black}{barrier} junction. \\
{\bf{\subsubsection{Uniformly gated graphene} }}
We apply RGFA to calculate the density of states (DOS) and conductance of graphene sheets of various widths ($W$). Fig. \ref {gnr}(a) shows the DOS (and conductance in inset) for $W$ = 10 and 100nm. Narrower nanoribbons show a gap as well as Van Hove singularities due to quantization for select chiralities while wider sheets show a quasi-linear bulk graphene density of states. In practice, nanoribbons also need attention to edge state dynamics, particularly the presence of strain and roughness \cite{tseng_09}.

A striking property of graphene is its anomalous integer quantum Hall effect. When the Fermi energy lies between two Landau levels (LL) in presence of a magnetic field, conventional two-dimensional electron gases show a vanishing longitudinal resistance $\rho_{xx} = 0$ and a quantized Hall conductance $G_H = 2q^2N/h$ where $N$ is a non-negative integer representing the number of filled Landau levels. In contrast, chiral quasiparticles in graphene exhibiting Berry phase show \cite{novoselov_05_nat, zhang_05} a Hall conductance $G_H = \frac{4q^2}{h}(N+1/2)$, with LL defined as $E_n = \frac{\hbar v_F}{r_C}sign(n)\sqrt{2|n|}$, where $r_C = \sqrt{\frac{\hbar}{qB}}$ is the cyclotron radius. This non-conventional sequence can be explained with the presence of a LL at E = 0 resulting in a four fold degeneracy at zero carrier density (from spin and valley degeneracy). On the contrary, Bilayer graphene LLs are defined as, $E_n = \hbar \omega_C\sqrt{n(n-1)}$, where $\omega_C$ is the cyclotron frequency. Now both $E_0$ and $E_1$ are at zero energy producing thereby an overall eightfold degeneracy at zero carrier density and QHE plateaus of $G_H = \frac{4q^2}{h}(N+1)$ \cite{mccann_06}. 

\begin{figure}
\centering
\includegraphics[width=2.2in]{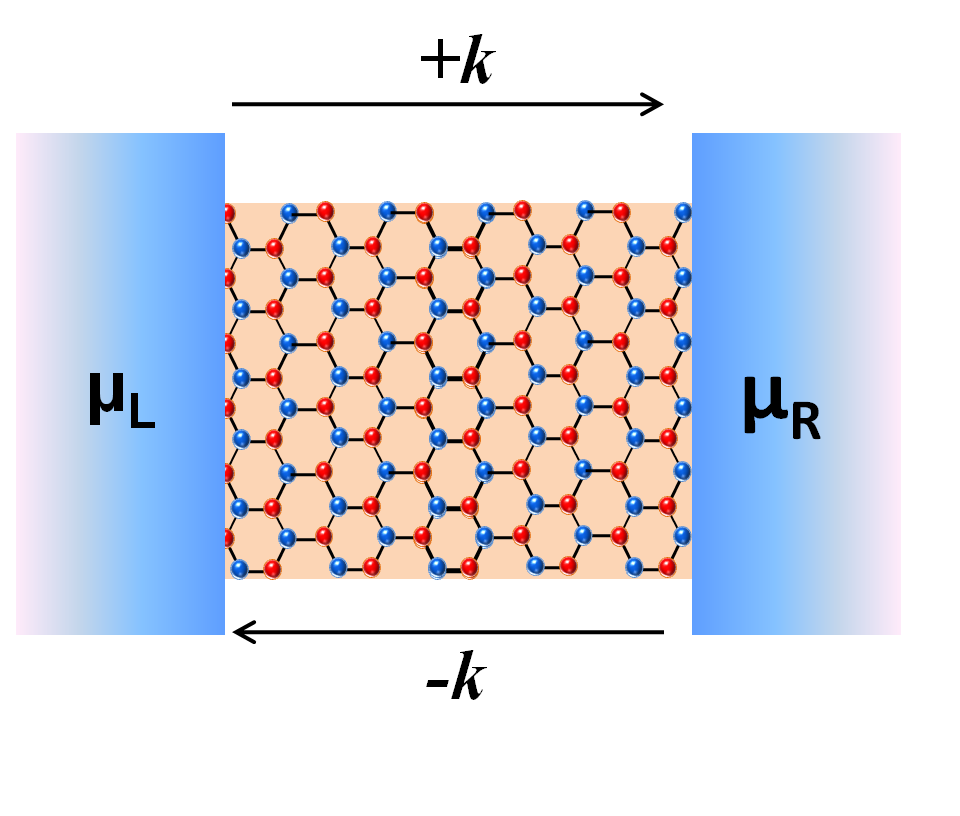}\quad
\subfigure{\includegraphics[width=3in]{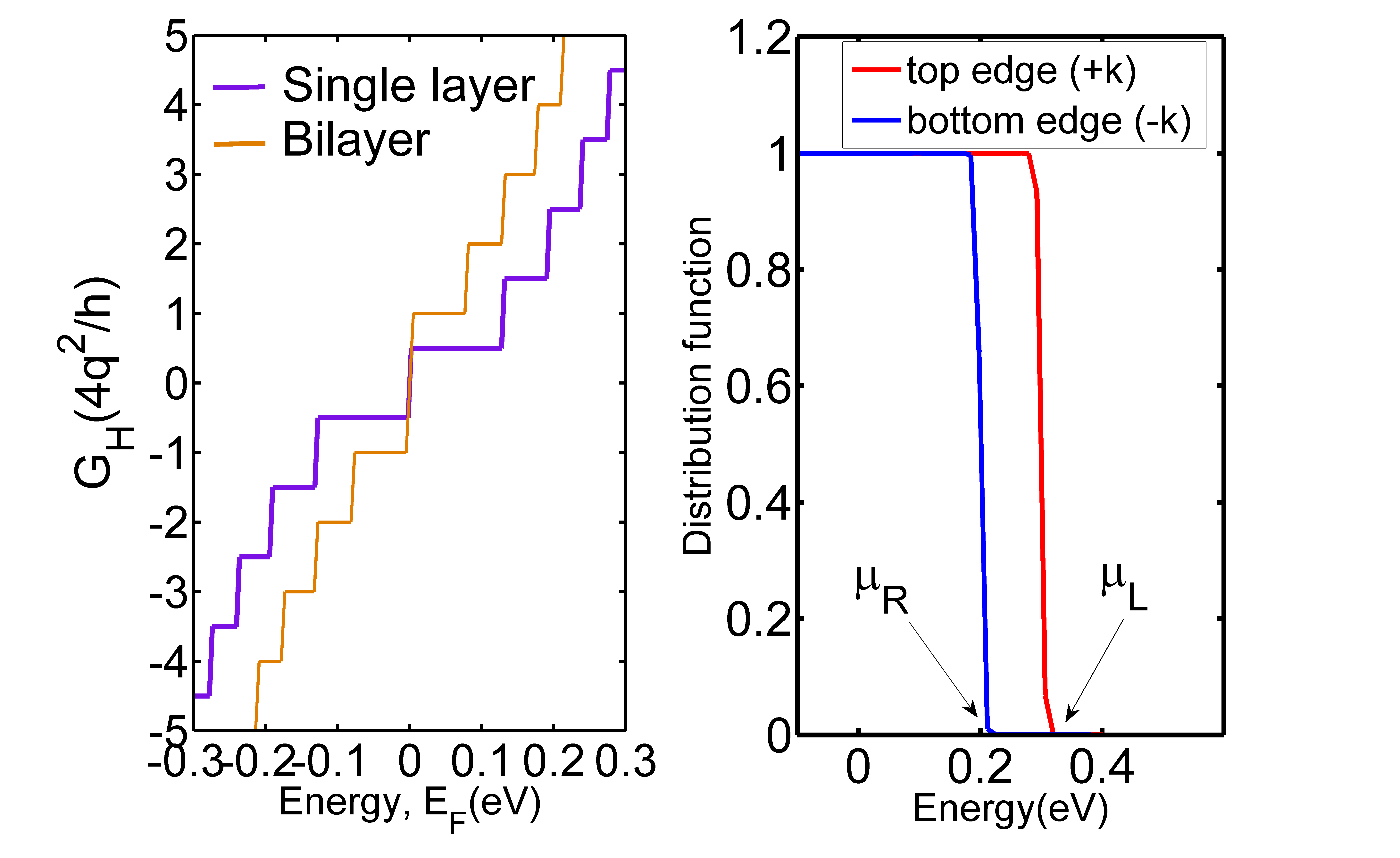}}
\caption{(a) In presence of a magnetic field, the current is carried by the edge states that separate into $+k$ states at the upper edge in equilibrium with the left contact and $-k$ states at the lower edge in equilibrium with the right contact. (b) The Hall voltage goes through plateaus at $\frac{4q^2}{h}(N+1/2)$ for single layer and $\frac{4q^2}{h}(N+1)$ for bilayer graphene with $N$ a non-negative integer. Note the presence of a jump at zero energy, which is not seen in a two-dimensional free electron gas (and arises from a half-filled Landau level at the Dirac point), and the additional factor of 2 in the plateau heights, arising from valley degeneracy. (c) For a placement of the Fermi energy between two Landau levels, electron distribution function at the top and bottom edges resemble \textcolor{black}{Fermi-Dirac distributions of the corresponding contacts.} Since each current carrying state sees a constant electrochemical potential {(along transport direction $x$)}, the longitudinal resistance vanishes.}
\label{qhe}
\end{figure}

\textcolor{black}{Numerical modeling magnetotransport in graphene requires a minor modification to the transport scheme outlined earlier. We replace the kinematic momentum with the quasi-momentum $\vec{k} \rightarrow \vec{k} - q\vec{A}/h$,  so that the plane wave terms in the Bloch representation pick up an additional phase of $e^{-iq/h \int \vec{A}\cdot \vec{dl}}$ where $A$ is the magnetic vector potential such that $\vec{\nabla} \times \vec{A} = \vec{B}$. Thus the hopping parameters between atoms `m' and `n' are now given by
\begin{equation}
t_{mn} = t_0\exp{\Biggl[i \frac{q}{h}\int_m^n\vec{A}\cdot\vec{dl}\Biggr]}
\end{equation}
For $z$ directed magnetic field, our guage is, $A = (-By,0,0)$. We can now turn on a magnetic field perpendicular to the sheet, modify the hopping terms as above, and extract the transvese (Hall) conductance as a function of varying Fermi energy location. The Hall conductance $G_H = I/V_H$, where $I = (2q/h)N(\mu_L-\mu_R)$ is the current, $N$ is the number of filled Landau levels and the Hall voltage $V_H = q\times$ \textcolor{black}{ the difference between the electrochemical potentials} of the $+k$ and $-k$ edge states. These states represent skipping orbits along the edges created by the cyclotron orbits in the bulk, and are separately in equilibrium with the left and right contact Fermi energies. }

Fig.~\ref{qhe}b shows the calculated Hall conductance yielding a series of plateaus for both single layer and bilayer graphene. Fig.~\ref{qhe}c shows the local carrier distribution functions $f(\pm \vec{k})$ obtained from the ratio of the local carrier density and the local density of states, in other words, the ratio 
\textcolor{black}{\begin{equation}
f(E) = \mathcal{G}^n_{L,L}(l,l)/i(\mathcal{G}_{L,L}(l,l)-\mathcal{G}_{L,L}(l,l))
\end{equation}}
 where $l$ is the index for an atom belonging to either top or bottom edge. The Hall voltage turns out to be $V_H = q(\mu_L - \mu_R)$ (Fig.~\ref{qhe}c) giving conductance plateaus for $G_H$. The electrochemical potentials ($\mu_L, \mu_R$) don't change along transport direction $x$ so that the longitudinal resistance (given by the drop in electrochemical potential along the channel) is zero.\\
{\bf{\subsubsection{Single p-n junction}}}
$pn$ junction heterostructures in graphene act qualitatively different from conventional $pn$ junctions. In normal semiconductors, the presence of a band-gap blocks electron flow, so that reducing (increasing) the built-in potential across a $pn$ junction with a drain bias leads to 
\textcolor{black}{an exponentially increased (reduced)} current, creating a rectifying current-voltage (I-V) characteristic. The lack of a bandgap removes such rectification in graphene $pn$ junctions (GPNJs). However, the chiral nature of the electron flow provides fascinating electron `optics' behavior, reminiscent of Snell's law, albeit with some notable twists. Across a $pn$ junction for instance, the conservation of transverse quasi-momentum ($k_y$) leads to anomalous Snell's law
\begin{equation}\label{snell}
K_{F1}sin\theta_1 = -K_{F2}sin\theta_2
\end{equation}
where the refractive index is set by the Fermi wavevector and thus the gate voltage applied. The opposite sign of the voltage across a $pn$ junction, arising when the quasimomentum component flips sign in going from conduction to valence band, means that the system acts 
like a negative index metamaterial.

\begin{figure*}
\centering
\includegraphics[width=5in]{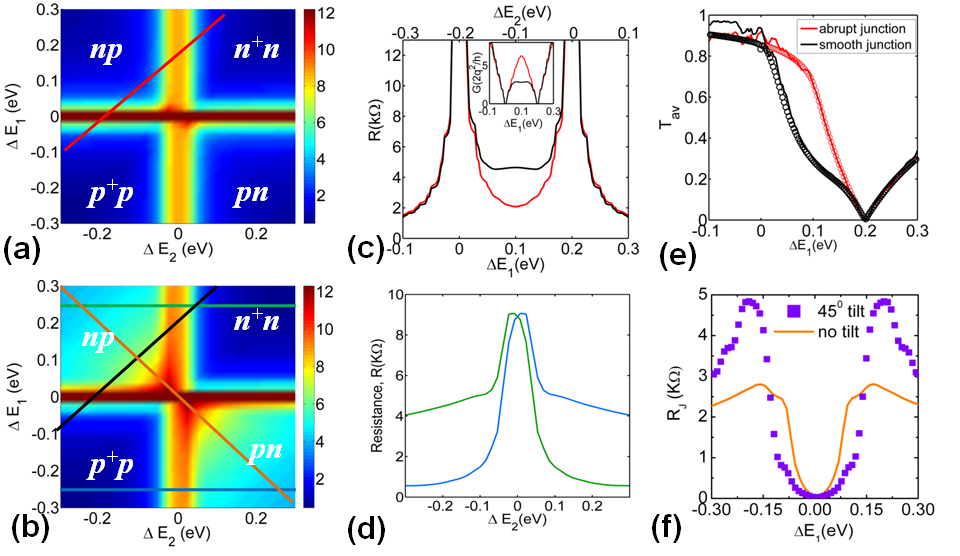}\quad
\caption{Numerical calculation (NEGF) of a single $pn$ junction conductance for both abrupt and smooth GPNJ. (a-b) Variation of total resistance with dopings in the two regions for abrupt (a) and smooth junction (c) Resistance as a function of Fermi energy ($E_F$) for a fixed built in potential $V_0$ = 0.2eV. The red (abrupt) and black (smooth junction) lines are resistance plots along the corresponding black and red lines in (a) and (b). Conductance (inset) which can be viewed as $G = G_0MT_{av}$ where $T_{av}$, pinches off at two points - one due to vanishing $M$ and at another one due to vanishing $T_{av}$, average transmission per mode. (d) The resistance asymmetry between $pn$ and $nn$ regime from plots along specific doping ($\Delta E_1$ lines from (b). (e) Extraction of $T_{av}$ numerically. The transmission of a symmetric GPNJ is 2/3 (at $E_F = 0$), which results in asymmetric conductance vs doping in GPNJ. The solid lines are from NEGF and the circles are from analytical calculation, Eq. \ref{uni}. (f) Junction resistance (Eq. \ref{eq_rj}) enhancement due to a tilt in the junction with respect to the tilt. This time we vary the doping along the diagonal (orange line in (b)) as done in the experiment \cite{sutar_12}.}
\label{cond1}
\end{figure*} 

While the electron trajectories are set by the above Snell's law, i.e., the conservation of transverse quasi-momentum, the actual transmission probabilities are set by the conservation of pseudospins. The pseudospins arise from the eigenvectors corresponding to the photon-like graphene E-k. The eigenfunctions are given by
\begin{equation}
\Psi = \Biggl(\begin{array}{c}1 \\ se^{\displaystyle i\theta}\end{array}\Biggr)e^{\displaystyle i(k_xx + k_yy)}
\end{equation}
where $\theta$ is the 2-D angle setting the quasi-momentum $\vec{k}$ and $s=\pm 1$ denotes the conduction or valence band \textcolor{black}{(An additional sign flip with respect to $k_x$ occurs near the other valley)}. Up and down pseudospin states are given by $\theta = 0$ and $\pi$, corresponding to the bonding and antibonding combinations of the $p_z$ dimer basis sets in graphene. Analogous to optics, we can derive the equivalent 'Fresnel equation' for transmission probabilities, by 
matching the pseudospinor components across the junction. Assuming a split distance `d' between the backgates defining the $pn$ junction, and a voltage difference $V_0$ between the gates, the transmission works out to be
\begin{eqnarray}\label{uni}
T(E_F,\theta_1) &=& \Biggl[\frac{cos\theta_1 cos\theta_2}{cos^2\biggl(\displaystyle\frac{\theta_1+\theta_2}{2}\biggr)}\Biggr]e^{\displaystyle -\pi {\hbar v_Fk_{F}^2}dsin^2\theta_1/V_0}
\label{Tklein}
\end{eqnarray} 
\textcolor{black}{below} a critical angle defined from the Snell's law (Eq. \ref{snell}), $\theta_1$ ($\theta_2$) representing the incident (refracted) angle for electrons. The first term shows that at normal incidence ($\theta_1  = 0$) the transmission function is unity regardless of the size of the barrier $V_0$. The absence of normal reflection is a manifestation of Klein tunneling, and arises because backscattering requires the flipping of pseudospins, in other words, a fast spatially varying potential. The second, exponential factor (drops out for unipolar $n^+n$ or $p^+p$ junctions) in the transmission
equation comes from tunneling across the slowly varying junction, where the transverse modes face a momentum-dependent band-gap, reminiscent of cut-off frequencies in a wave-guide.  

The overall conductance of a GPNJ can be calculated by summing over all transmissions, $G = G_0\sum_{\theta}^{M}T(\theta)$, which can be thought of as the total number of modes times an average transmission. Thus the average transmission over all modes at a particular energy is given by $T_{av} = {G}/{G_0M}$, where number of modes, $M= {4|E_F|}W/{\pi\hbar v_F}$. From RGFA calculation, we get the total conductance $G = Tr(\Gamma_1\mathcal{G}\Gamma_2\mathcal{G}^{\dagger})$ in units of $G_0 = 2q^2/h$. 

\begin{figure*}
\centering
\includegraphics[width=3.5in]{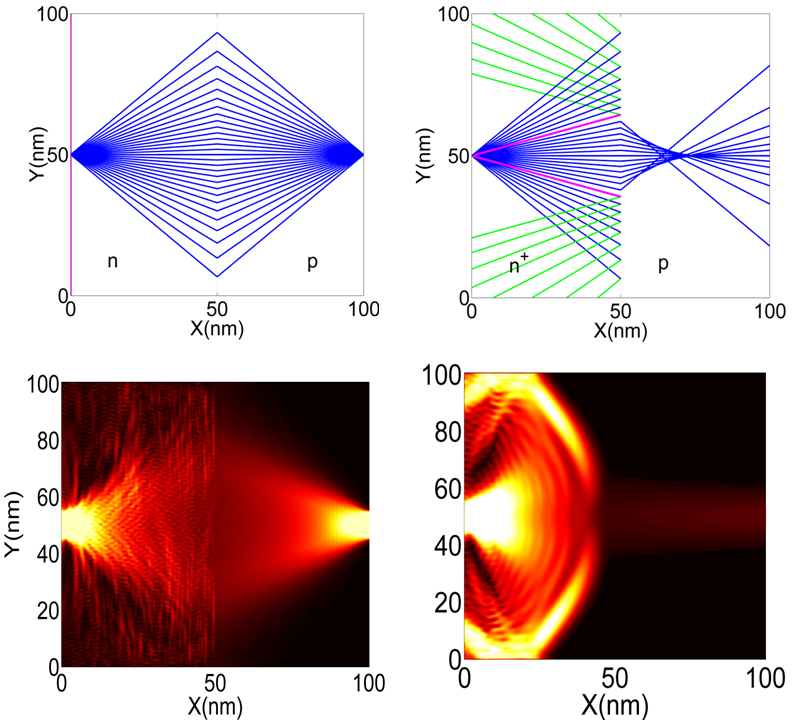}\quad
\caption{Tracking electron trajectories in GPNJ, negative index makes the refraction angle negative - resulting in spatial focusing of current flow for equal dopings (left column). Right column shows total internal reflection when the refracted side has lower doping, thus reflecting electrons outside the critical angle.}
\label{veselago}
\end{figure*}

Figs.~\ref {cond1} (a,b) show the doping-dependent resistance of a GPNJ for a 100nm wide graphene sheet, for abrupt and slowly varying junctions respectively. In fig. \ref{cond1}, we plot the variables against the shifts in the Dirac point in both regions, i.e. $\Delta E_1$ and $\Delta E_2$. The resistances at the upper left and lower right corners of the plot are higher than the other two ($pn$ vs uniformly doped), resulting in  an asymmetric resistance vs doping in fig. \ref {cond1}(d) (plotted for specific doping values $\Delta E_1$ as indicated with \textcolor{black}{horizontal lines in \ref{cond1}(b)}). 

The WKB term in Eq. \ref {uni} is only present in the $pn$ junction regime, and that is why only the $pn$ junction resistance is affected while going from abrupt to smooth junctions (a-b). For a $pp$ or $nn$ junction the Fermi energy does not cross \textcolor{black}{the smoothly varying Dirac point anywhere in the device and the transmission expression only includes the wave-function mismatch term}. Fig. \ref{cond1}(c) shows the resistance variation for a fixed built-in potential $V_0 = \Delta E_1-\Delta E_2$ (along black and red lines in fig. \ref{cond1}(a) and (b)). In the conductance variation (inset), we see effectively two Dirac points, which has a simple physical explanation. Recall that the normalized conductance can be decoupled into the \textcolor{black}{mode count} from one end times the average transmission over to the other side
\begin{equation}
G/G_0 = M_1T_{12} = M_2T_{21}
\end{equation}

The left conductance minimum at $E_F = -0.1eV$ is where $M_1$ becomes zero, while the right one at $E_F = +0.1eV$ is where the average transmission for all modes, $T_{12}$, becomes zero. To make this clear, we calculate $T_{av}$ numerically (Fig. \ref {cond1}(e)), clearly showing a vanishing transmission at the second Dirac point. To calculate $T_{av}$, we first simulate a graphene device with uniform doping and extract the overall mode count from the ballistic conductance ($G_1 = MG_0$). We then simulate the device with different dopings (finite built-in potential, $V_0$, and conductance $G_2 = G_0M T_{av}$). The ratio of $G_2(E_F)$ and $G_1(E_F)$ at each energy gives us $T_{av}(E_F)$. For example, for a symmetric GPNJ (at $E_F = 0$), we have
\begin{eqnarray}
G/G_0 &=&\sum_{\theta} T(\theta) \approx \int_{-\pi/2}^{\pi/2} \frac{T(\theta)}{\Delta \theta}d\theta\nonumber\\
&=&\int_{-\pi/2}^{\pi/2} \frac{cos^2\theta}{\Delta k_y}k_Fcos\theta d\theta = \frac{Wk_F}{2\pi}\int_{-\pi/2}^{\pi/2} cos^3\theta d\theta\nonumber\\
&=&\frac{2}{3}M(E)
\end{eqnarray} leading to $T_{av} = 2/3$, which is close to what we get in the numerical calculation. Thus the conductance of a symmetric GPNJ is 2/3 of the uniformly doped graphene sheet with the same Fermi energy ($E_F$). In {{\textcolor{black}{Fig.~\ref {cond1}(e)}}} we also show the analytical calculation of $T_{av}$ (circles) in the same manner as done above but this time using the transmission in Eq. \ref{uni}. The analytical calculation matches very closely with the numerical calculation (solid lines).  

A more direct evidence of chiral tunneling is seen from the resistance across a tilted junction. We use the following equation to extract the junction resistance \cite {datta_97}, that eliminates the quantized resistance from the contacts.
\begin{eqnarray}\label{eq_rj}
R_J = (\frac{4q^2}{h})^{-1}[\frac{1-T_{av}}{MT_{av}}]
\end{eqnarray} We calculate $M$ and $T_{av}$ numerically from NEGF as discribed earlier. Fig. \ref{cond1}f shows the variation of $R_J$ plotted against $\Delta E_1$ (along orange line) for a 100nm wide graphene sheet. For low carrier density \textcolor{black}{in the middle}, the background doping ($n_0$) makes the device an $n^+n$ or $p^+p$ junction and $R_J$ is low. For $n_{1,2}>n_0$, \textcolor{black}{i.e., in the wings of the plot}, the device reaches the $p-n$ junction regime and $R_J$ becomes high. \textcolor{black}{This characteristic} doping dependence of junction resistance has been seen in experiments in the past \cite{huard_07,stander_09}, \textcolor{black}{usually captured by the odd resistance}, $2R_{odd} = R(n_1,n_2)-R(n_1,-n_2)$. \textcolor{black}{The chiral tunneling can be seen more directly by varying the tilt angle, as seen in experiments recently \cite{sajjad_12}}. The purple squares show the trend for a tilted GPNJ, where a pronounced peak is seen in $R_J$. With \textcolor{black}{increasing} tilt \textcolor{black}{ between the junction and the transport axis normal to the contacts}, the incident angles\textcolor{black}{\bf{s}} of the incoming modes are effectively increased, leading to higher resistance. Such an increase in junction resistance is seen in a recent experiment \cite{sutar_12} and a detailed theoretical treatement can be found in \cite{sajjad_12}. \textcolor{black}{Interestingly, the results suggest the absence of specular reflection from the edges of the graphene sample.}

We will next apply the current density formalism in RGFA to show the electron trajectories inside the GPNJ device. The total current at an atomic site is equal to vector sum of all current components to nearest neighbor atoms (from Eq. \ref{current_den}). The classical ray tracing analysis from Snell's law is shown in Fig. \ref {veselago}(a). For equal dopings on both sides ($k_{F2}=k_{F1}$), the angle of refraction is exactly equal (with a negative sign) to the incident angle. As a result, a group of electrons originating from a point contact  focus back to one point on the refracted side \cite{ref1}. Fig. \ref {veselago}(b) shows \textcolor{black}{the trajectories} when the doping at the refracted side is smaller than that at the incident side ($k_{F2}<k_{F1}$) making the critical angle smaller than $\pi/2$. \textcolor{black}{The geometrical `optics' trajectories corresponding to electron focusing and total internal reflection} are \textcolor{black}{reproduced by the calculated current density withing an atomistic} RGFA simulation,  shown in Fig. \ref {veselago}(c), (d). \textcolor{black}{A small source contact (10nm wide) is placed $50$nm to the left of the pn junction, and electrons are injected with a small drain bias ($V_D$ = 0.08 V) around a Fermi energy. When the gates are biased symmetrically around the junction, although the electrons see a voltage bias along the drain that spans the entire device width, the $pn$ junction Hamiltonian and associated pseudospin conservation causes the electrons to focus to a small point at the drain. For an n$^+$p junction, the electrons incident above the critical angle are unable to preserve their transverse quasi-momentum and reflect back, while those within critical angle tunnel to the opposite band on the other side.} 

We now investigate the quantum Hall plateaus for a single $pn$ junction. According to recent experiment \cite{williams_07}, the conductance in the $pn$ junction
regime follows the Ohmic conductance rule
\begin{eqnarray}
G(q^2/h) = \frac{\nu_n\nu_p}{\nu_n+\nu_p};\,\,\,\,\,\,\,\,\, Ohmic\,\, plateaus
\end{eqnarray}$\nu_n$ and $\nu_p$ are the filling factors in the two segments of the device. This was explained by Abanin $\textit{et. al.}$ \cite{abanin_07} in terms of mode mixing at the $p-n$ interface for large systems with diffusive transport. As graphene filling factors are $\left\{2,6,10,...\right\}$, it produces $pn$ junction plateaus as $\left\{1,\frac{3}{2},...etc.\right\}$ for filling factor combinations of $(2,2)$, $(2,6)$ etc.  matching closely with the experiment \cite{williams_07}. 

This result changes considerably for coherent ballistic transport for smaller structures. Tworzydlo $\textit{et. al.}$ \cite{twor_07} showed that for armchair nanoribbons the conductance plateaus become
\begin{eqnarray}
G = \frac{1}{2}G_0(1-cos\Phi); \,\,\,\,\,\,\,\,\, Ballistic\,\, plateaus
\end{eqnarray} independent of individual filling factors $\nu_n$, $\nu_p$. $G_0=2q^2/h$ is the lowest Hall plateau for graphene. $\Phi$ 
is the angle between valley isospins at the top and bottom edges and depends on the chirality of the nanoribbon. Depending on the number of hexagons $N$ along transverse direction, $\Phi=\pi$ for $N=3M+1$ and $\Phi=\pi/3$ otherwise. This leads to 
\begin{equation}
G(q^2/h) = \left\{ \begin{array}{ll}
2 \,\,\,\,\,\,\,\,\,\,\,\,  & if \,\,\, \textrm{N = 3M+1}\\
\frac{1}{2}    & \textrm{otherwise}\end{array} \right.
\end{equation}The ballistic to ohmic cross over for the plateaus can be recovered by using interface and edge disorder \cite{low_qhe_09}. 

The magnetotransport for zigzag ribbons cannot be explained with valley isospin arguments. Akhmerov $\textit{et. al.}$ \cite{akhmerov_08} proposed the theory of valley valve effect with intervalley scattering which leads to complete suppression of conductance in the zigzag $pn$ junction, if the number of atoms across the ribbon is even. 
\begin{figure}
\centering
\includegraphics[width=3.4in]{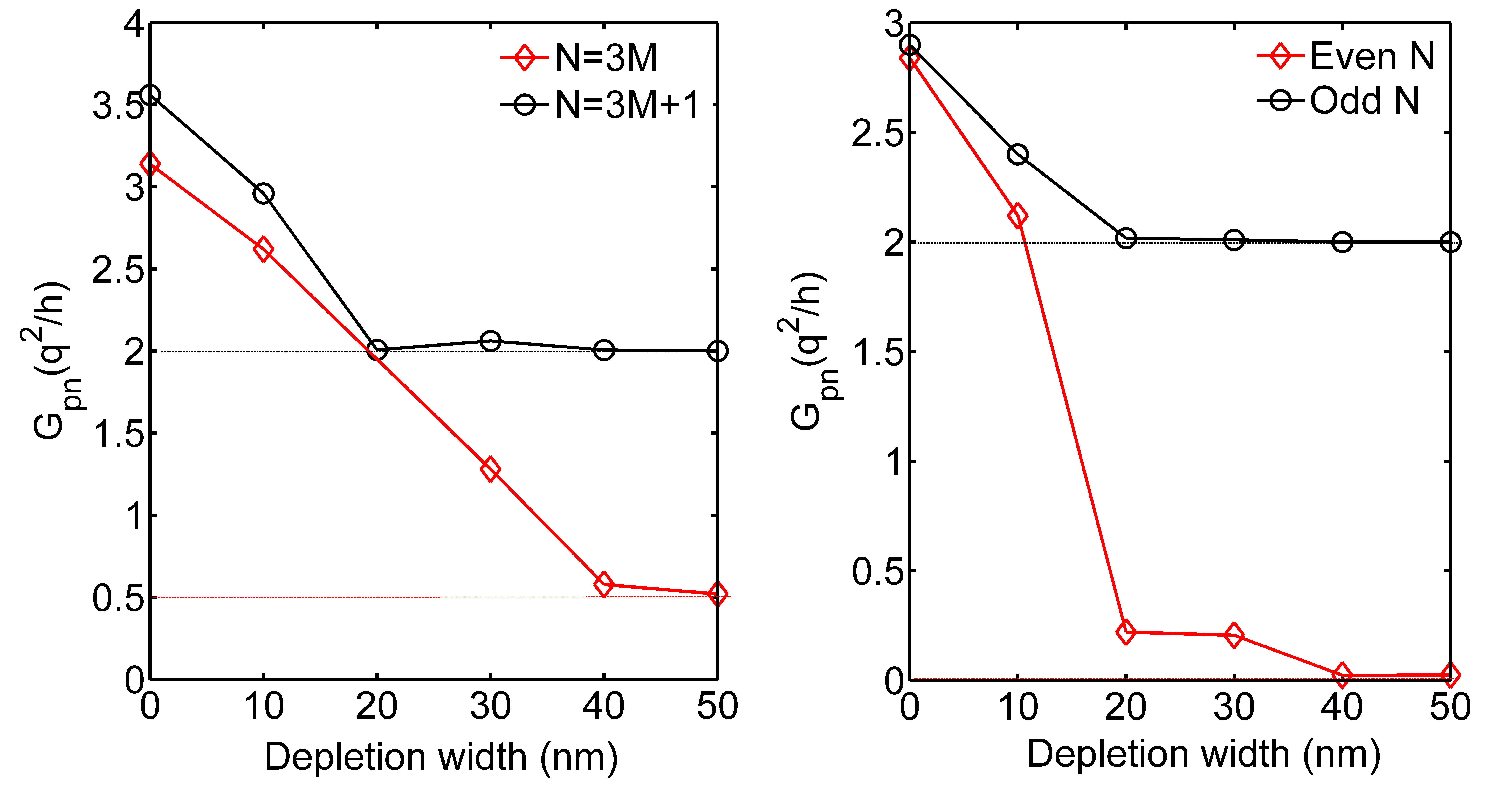}\quad
\caption{Magnetotransport in graphene $pn$ junction. Magnetic field, B=10T and filling factors are $\nu_n = \nu_p = 6$.  (Left) Change of conductance with depletion width, $d$ for Armchair ribbons. For high depletion width, we get the expected plateaus of 2 and $\frac{1}{2}$. (Right) Plateaus of 0 and 2 for Zigzag ribbons. Dotted horizontal lines show analytical predictions.}\label{qhe_pn}
\end{figure}
\begin{equation}
G(q^2/h) = \left\{ \begin{array}{ll}
0\,\,\,\,\,\,\,\,\,\,\,\,    & if \,\,\,\textrm{N = 2M}\\
2    & \textrm{otherwise}\end{array} \right.
\end{equation}

Impact of depletion width $d$ (spacing between two gates) on GPNJ ballistic magneto-conductance for both armchair and zigzag ribbons are shown in Fig. \ref{qhe_pn}. Magnetic field B = 10T and dopings are $\Delta E_1 = \Delta E_2 = 0.15$eV producing $\nu_n = \nu_p = 6$. Conductance values approach analytically predicted numbers only when the depletion width is sufficiently large ($\textgreater$ 30nm), filtering out higher Landau Levels (LL) \cite{low_qhe_09}. It requires larger depletion width as individual filling factors get higher. In most cases, the lower plateaus ($\frac{1}{2}$ and 0 for armchair and zigzag respectively) are more difficult to achieve than the higher one (2).\\

\begin{figure}
\centering
\includegraphics[width=3.4in]{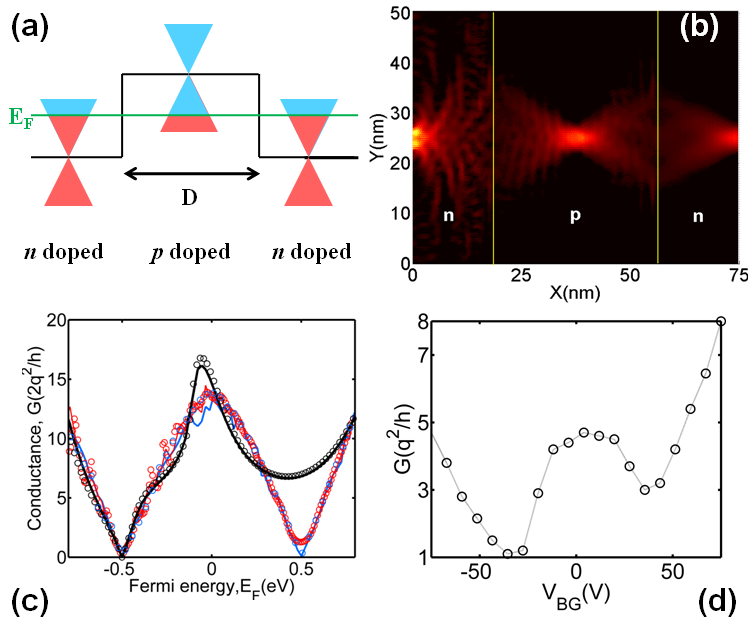}\quad
\caption{Physics of $npn$ junction. (a) Band diagram. The black line shows the change of Dirac point as a result of different dopings at different portions of the device. Blue (red) region in the E-k indicates empty (filled) states. (b) Electron trajectories in such junctions, multiple focusing takes place, (c) Ballistic conductance as a function of Fermi energy for a fixed barrier height for different barrier width, $D$=5, 25 and 100nm (black, red and blue) . Width of the device is W = 50nm. The solid lines are from numerical NEGF calculation, while the circles are from analytical calculation, Eq. \ref{tr_kats}. (d) Experimental data from Ref. \cite{pkim_07}. Conductance in the experiment is much lower due to scattering mechanisms.}
\label{graphene_npn}
\end{figure}

{\bf{\subsubsection{n-p-n junction}}}
By applying a step like potential, we \textcolor{black}{can realize an} $npn$ junction. The dopings can be done by either a combination of global back gate and top gate \cite{stander_09} or by selective chemical doping. In fig. \ref{graphene_npn}(a), we show the energy band diagram for the device. We consider both junctions as abrupt, $d = 0$, while the extent of the $p$ type barrier region is $D$. In fig. \ref{graphene_npn}(c), we show an NEGF calculation of  the npn conductance for various $D$ values. For larger $D$, two Dirac points are evident from the pinched off conductance, but for smaller $D$s we see a considerable tunneling around the second Dirac point ($E_F$ = 0.5eV), where the critical angle for the incident electrons is supposed to be very small. The first dip happens at the Dirac point of the incident n region where $M_1 = 0$ as before, while the second corresponds to the electrons in the n region aligned with the Dirac point of the barrier p region and the transmission $T_{12}$ is small. In a single $p-n$ junction, the modes with higher angles than the critical angle reflect back \textcolor{black}{as they do not have any propagating states to tunnel into while preserving their transverse quasimomentum}. But in the $npn$ junction case, the length of the forbidden region $D$ is \textcolor{black}{finite, and the electrons even while aligned with the Dirac point of the central p region can still preserve their quasi-momenta by  tunneling to the other side}. This increased conductance has been seen experimentally \cite{pkim_07} in the past \textcolor{black}{(Fig.~\ref{graphene_npn})}. In the limit when $D$ is very large, we approach the single $p-n$ junction case \textcolor{black}{we worked out in Fig. \ref{cond1}(c) inset}. And when $D$ is very small, we approach the uniform doping ($n$ doped here) case \textcolor{black}{through significant tunneling}. Note that we \textcolor{black}{also see} oscillations for $npn$ junctions, which originate from the Fabry Perot cavity formed by the electrostatic barrier \cite{pkim_07,young_09}. Both the tunneling and \textcolor{black}{the resonant} oscillations can be captured analytically \textcolor{black}{by matching eigenvector components across the pnp junction, as} worked out in Ref. \cite{katsnelson_06}.
\begin{equation}\label{tr_kats}
r = \frac{2ie^{i\phi}sin(q_xD) \times (sin\phi -ss'sin\theta)}{ss'[e^{-iq_xD}cos(\phi+\theta)+e^{iq_xD}cos(\phi-\theta)]-2isin(q_xD)}
\end{equation}
\textcolor{black}{giving us a} transmission, $T(\phi) = 1-R = 1-|r|^2$. $q_x$ is the wave vector inside the barrier, $q_x = \sqrt{(E_F-V_0)^2/(\hbar^2v_F^2)-k_y^2}$, $\theta = tan^{-1}(k_y/q_x)$ and $s=sign E_F$ and $s'=sign(E_F-V_0)$. We then sum over different modes, set by the width of the graphene sheet, to get the total conductance $G(E_F)$ in units of $G_0 = 2q^2/h$ (Fig. \ref{graphene_npn} (c), ,circles), \textcolor{black}{in excellent agreement with the NEGF atomistic calculation and qualitatively  with experiments (Fig.~\ref{graphene_npn}))}.\\

{\bf{\subsection{KSF to simulate bulk graphene and GPNJ}}}

\textcolor{black}{The inset in} Fig. \ref {gnr}(b) shows the calculated conductance of bulk graphene from KSF. Note that the peaks due to finite width quantization in Fig.~\ref{gnr} are \textcolor{black}{removed by} the application of periodic boundary conditions. The black circles show the trend that is expected from a linear approximation of the number of modes $M = {2W|E_F|}/{\pi\hbar v_F}$ at low energy. The main figure in Fig. \ref{gnr}(b) shows the bulk graphene density of states calculated from three methods. The red line shows the result from an \textcolor{black}{atomistic} NEGF calculation. The black circles are from an analytical formula \cite{neto_09}, while the blue line is from a numerical extraction of \textcolor{black}{the density of states (DOS)} from the graphene $E-K$ relation. All three calculations match very well with one another.\\

\begin{figure}
\centering
\includegraphics[width=3.2in]{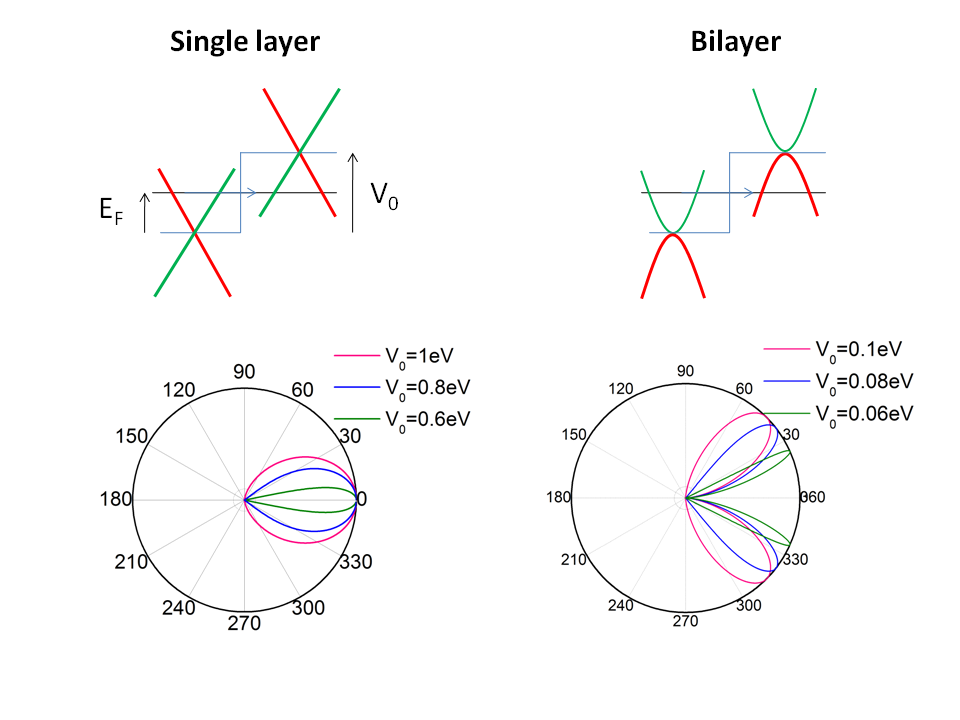}
\caption{Angle resolved transmission in single layer and bilayer graphene for a fixed Fermi energy and for various built-in potential $V_0$. At left ($E_F = 0.5$eV), Klein tunneling with maximum transmission at normal incidence. At right ($E_F = 0.05$eV), Klein reflection in bilayer graphene for normal incidence. Top pannel shows the forward and backward moving branches colored according to the pseudospin texture for normal incidence.}
\label{chiral}
\end{figure}

KSF approach also allows us to show the angle (mode) dependent chiral tunneling in graphene. We use Eq. \ref{emode} to get $T_{k_y}$ which is converted into a polar plot (Fig. \ref{chiral}) using $\theta = sin^{-1}\frac{k_y}{k_F}$, for transmission per spin per valley. For single layer graphene the transmission becomes unity for normal incidence (known as Klein tunneling). On the contrary, for bilayer graphene, the reflection becomes unity (Klein reflection). The tranmission ultimately depends on how well the wave-functions match on both sides of the junction (branches with same pseudospin components have same color in Fig. \ref{chiral}). \\

{\bf{\subsection{Combining KSF and RGFA}}} 
We now combine the two transport formalisms
KSF and RGFA to a graphene $p-n$ junction and show how this method allows us to 
simulate  very long, micron-sized devices with periodic boundary conditions along 
the transverse direction. \textcolor{black}{We apply this technique to numerically reproduce the tunnel
transmission across a GPNJ with split gates, which requires simulating a long device to extract the proper decay constant. We assume that the Dirac point varies linearly
across the junction for simplicity (Fig.~\ref{cond1}(a)) 
In presence of a slowly varying potential across the GPNJ, the}  transverse modes with finite wave-vector $k_y$ \textcolor{black}{act like waveguides with a cut-off frequency and thus need} to overcome a real tunnel barrier. 
A smooth GPNJ thus selectively transmits the low angle electrons \cite{falko_06}, \textcolor{black}{with suppression of the higher angle modes with increasing length of the junction transition region}. The fundamental normal mode still does not see a barrier and thus transmits perfectly, \textcolor{black}{reflection once again eliminated by a pseudospin dominated selection rule}, retaining the Klein tunneling properties of GPNJ. 

\begin{figure}
\centering
\includegraphics[width=3in]{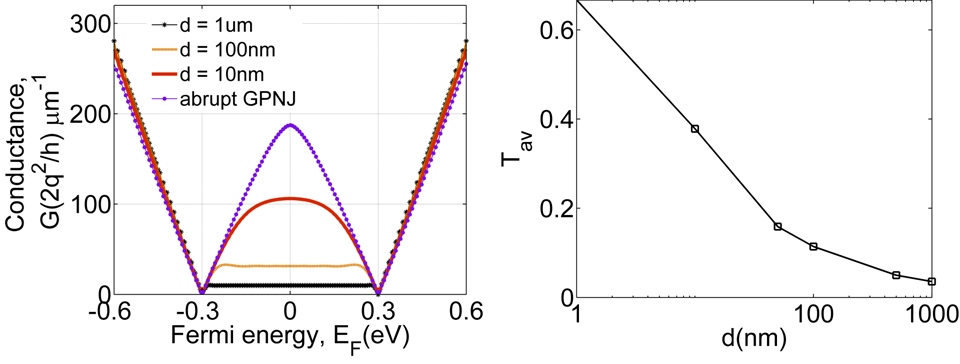}
\caption{KSF-RGFA simulation of smooth GPNJ. (a) Conductance characteristics as a function of Fermi energy for a fixed built-in potential $V_0$ = 0.6eV, conductance is suppressed in the $pn$ junction regime due to the exponentially decaying transmission of higher order modes. Inset: Average transmission per mode $T_{av}$ for a symmetric GPNJ ($E_F = 0$ point in (a)) vs the separation between two gates.}
\label{gpnj1}
\end{figure} 

The conductance of a symmetric (equal \textcolor{black}{and opposite} doping on both sides) smooth GPNJ, \textcolor{black}{using the expression from Eq.~\ref{Tklein}}, becomes
\begin{eqnarray}
G &\approx & \frac{4q^2}{h}\int_{-\theta_0}^{\theta_0} \frac{d\theta}{\Delta\theta}e^{-\pi k_Fd\theta^2} \approx \frac{4q^2}{\pi h}\sqrt{\frac{k_F}{d}}W
\end{eqnarray} where $\theta_0 = (\pi k_Fd)^{-1}$ is the effective critical angle imposed by the smooth junction \cite{falko_06} and $\Delta\theta = \Delta k_y/(kcos\theta)$ is the angular separation between adjacent modes. 
\textcolor{black}{Numerically capturing this exponential decrease in conductance with gate split distance $d$ requires simulating a long device}. The definitions of the unit cell Hamiltonian are now changed by the following. 
\begin{eqnarray}
H_{L,k} = H_L+t_{L,L+1}e^{ika}+t_{L,L+1}^{\dagger}e^{-ika}
\end{eqnarray}where {\bf{\textcolor{black}{$t_{L,L+1}$}}} is the layer to layer coupling matrix in the transverse direction. This will describe a layer Hamiltonian with periodic boundary condition. Similarly we use $\alpha_{k_y}$ to calculate $g_{k_y}$ as described in section II. 
The $T_{av}$ vs $d$ is shown in inset of Fig. \ref{gpnj1} with $d$ extended in the micron regime. \\

{\bf{\section{Conclusions - the role of numerical simulation}}}
\textcolor{black}{Graphene and its heterojunctions constitute a fascinating system that demonstrate rich physics such as anomalous quantum Hall, chiral tunneling, metamaterial behavior and pseudospintronics. Extended to other sub-systems such as bilayer graphene, this richness proliferates with Lifschitz transitions, anti-Klein tunneling, more anomalous Hall signatures and so on. The simplicity of the low-energy 2-D bandstructures of graphene and its progeny allows pen and paper demonstration of these physical concepts. The NEGF approach provides a common computational framework to verify these concepts atomistically, whereby a small modification to the Hamiltonian (e.g. a barrier or a field-induced phase in the coupling parameter) can manifest these physical concepts with very little effort, even in tougher geometries such as multiple junctions and tilted heterostructures. When comparing with experiments, however, one needs to worry about the sensitivity of these effects to scattering, edge states, quantization, smooth junctions, strain, roughness and so on. This is where computation plays a definitive role. It was hard to anticipate {\it{a-priori}} what the effect of specular edge scattering would be on the resistance of a tilted junction, but we now know through numerical simulations \cite{sajjad_12} that it creates a sweet-spot in the scattering profile that experiments do not show. Such `smoking-gun' demonstrations allow us to judge the feasibility of devices that utilize unconventional electronic switching in graphene, such as the creation of transmission gaps with subthermal switching \cite{sajjad_11}. We can now atomistically simulate graphene systems whose sizes are comparable to experimental dimensions, and directly demonstrate many of the physical effects that such devices are based on in presence of non-idealities. Further extensions will involve the role of self-consistent screening (Poisson), charge puddles, defects and incoherent scattering from remote optical phonons.}

\begin{acknowledgements}
The authors acknowledge financial support from INDEX-NRI. Redwan Sajjad would like to thank Khairul Alam and Golam Rabbani for the assistance in implementing RGFA. Authors also thank Eugene Kolomeisky, Branislav Nikolic, Farzad Mahfouzi for useful discussions. 
\end{acknowledgements}

\bibliographystyle{IEEEtran}

\end{document}